**Non-coding DNA programs express adaptation and its universal law**

**Mark Ya. Azbel'**

**School of Physics and Astronomy, Tel-Aviv University,**

**Ramat Aviv, 69978 Tel Aviv, Israel**

**Summary**

Significant fraction (98.5% in humans) of most animal genomes is non- coding "dark matter". Its largely unknown function (1-5) is related to programming (rather than to spontaneous mutations) of accurate adaptation to rapidly changing environment. Programmed adaptation to the same universal law for non-competing animals from anaerobic yeast to human is revealed in the study of their extensively quantified mortality (6-21). Adaptation of animals with removed non-coding DNA fractions may specify their contribution to genomic programming. Emergence of new adaptation programs and their (non-Mendelian) heredity may be studied in antibiotic "mini-extinctions" (22-24). On a large evolutionary scale rapid universal adaptation was vital for survival, and evolved, in otherwise lethal for diverse species major mass extinctions (25-28). Evolutionary and experimental data corroborate these conclusions (6-21, 29-32). Universal law implies certain biological universality of diverse species, thus quantifies applicability of animal models to humans). Genomic adaptation programming calls for unusual approach to its study and implies unanticipated perspectives, in particular, directed biological changes.

## Introduction

Unusual approach to genomics (via analysis of readily available mortality data) unravels unanticipated function of non- coding DNA.

Mendel study of hereditary traits, later related to few alleles, revealed units of heredity. Such traits are rare, and Mendel laws were disregarded until rediscovered. Arguably, similar bias in mortality studies significantly delayed discovery of non-genetic heredity. In 1825 Gompertz (33) started ongoing search (34-38) for the law of universal mortality. Mortality of evolutionary unprecedented human and laboratory animals, which are mostly protected (further protected populations) from competition with other animals, is extensively quantified. Their mean lifespan (immature stages including) exhibits extraordinary phylogenetic irregularity (6-21). Immature nymph stage in mayfly and cicada Magicicada is up to 4 and 23 times longer than embryo stage in humans, 100 and 1,700 times longer than larvae stage in Drosophilae. Mature Mean Lifespan (MLS) is about 1-2 days in single cell yeast - and mayfly whose adults do not eat, rapidly senesce and die after mating; 20-50 days in nematode - and Drosophilae. MLS of human is closer to hydra with no signs of aging for 4 years, and possibly immortal (39), than to mice with MLS~1 year. From mayfly to humans, MLS increases ~30,000 times, the ratio of immature to mature time decreases more than 100,000 times. Certain mutations change MLS in mice 1.6-fold (8), and in nematode 3.6 times (17). Human MLS significantly and irregularly changes with calendar year. For instance, female MLS (calculated according to mortalities at different ages in a given calendar year- further only such data are considered) in Sweden was 18.8; 41.4; 28.9; 47.3 years in 1773; 1774; 1809; 1823 correspondingly. Mortality depends on genotypes, phenotypes and their heterogeneity, environment, living conditions and their change, life histories, etc- see, e.g., entire issue of Cell 120, #4 (2005). The Gompertz law did not quantify all these factors, was often very inaccurate, thus disregarded in all theories of aging and mortality (40). To estimate and forecast human mortality, demographers developed over 15 different approximations (41). The lifespan of four populations of inbred 3X3 male drosophilae in presumably identical shell vials varied from 18.6 to 34.3 days (9). Thus, living conditions, which look as micro-environmental variations, may in fact be very different and age dependent.

However, raw data demonstrate that when a single number relates age of a given species to human age, then under certain conditions survivability dependence on such scaled age and MLS is predominantly the same for different populations of species as diverse as anaerobic and aerobic yeast, nematode, mayfly, drosophilae, mouse, and human. Such invariance to all other factors is sufficiently restrictive to yield the variables, exact formula, and conditions of validity of the universal mortality law (which in special cases reduces to the set of Gompertz laws) in a general case. The law is verified with all available data; deviations from the law quantify living conditions of a given populations and species. In rapidly changing environment, mortality accurately adjusts to the law. Universality of the law and adjustment to it in diverse species, as well as over two centuries of human data, proves that such law and the *possibility* of such adjustment are hereditary. Thus, while such adjustment is acquired and rapidly reversible, its universality is provided by unusual express adaptation. Hereditary exact law and express adaptation to it must be programmed (rather than related to specific coding genes and slow spontaneous mutations) by non-coding DNA.

For the sake of biologists, all formulas are shifted to the last section

**Results**

In mature stage probability $\ell$ to survive to the age x, as well as MLS e for each species, sex and population (country and calendar year of birth and death for humans) are listed in readily available life tables (6- 21). Figure 1a presents the population fraction $\ell^*(R, e)$, which, according to raw data, survives to any relative age R=x/e, for anaerobic yeast [(20), e=15.75 generations~1 day], nematode [(12), e=36 days], human [(6), e$\approx$83 years]. Despite drastic difference in biology and 30,000 times change in MLS, their $\ell^*(R, e)$ curves are nearly the same. (Stochastic deviations are significant when, e.g., there are only 4 or less nematodes survivors). These curves are similarly close to those of aerobic yeast [( 42), e=21.6 generations], mayfly (11), drosophilae (10), mouse (8), correspondingly e=1.6, 36, 714 days. The shapes of human curves with e$\approx$64, 42, 32, 20 years, are very different, yet close for close values of their e. Scale "biological time" of different species according to the values of e in their close curves. This equates 1 generation of anaerobic and aerobic yeast; 1 day of nematode, mayfly, drosophilae and mouse to correspondingly F=5.5 and 4; 2.4, 50.7, 1.8, 0.17 human years. For their different populations Fig. 1b manifests predominantly universal survivability $\ell(X, E)$ to any scaled age X= Fx for the scaled mean lifespans (SLS) E= Fe $\approx$ 20, 32, 42, 64, 83, 130, 295 (here and on X and E in years). Mortality is stochastic, thus significant fluctuations in survival curves of small animal populations [21, 26 mice (8); 48 (12), 68, 39 (17) nematodes; yeast 16,15,14 anaerobic (20) and 35, 46, 45 aerobic (42, 15, 16) initial yeast cells; at, e.g., E=295 years (17) on average one nematode dies per 7 years] and smooth curves of large human populations. All populations but drosophilae are heterogeneous. Figure 1a implies (here and on all details and derivation of accurate formulas see in Methods) that survivability dependence on the SLS E is universal and piecewise linear, when  SLS in all groups of any population are restricted to a single universal interval. Denote the latter situation as "restricted heterogeneity". (A general case of populations with unrestricted heterogeneity yields somewhat more complicated universal law).  This agrees with the survivability dependence on SLS E in the considered species- see Fig. 1c for X=30 and 85. When, e.g., X=85, linear slopes at the intersections increase 2.4 times at $E_1$=43, then 4.3 and 3.4 times at $E_2$ =62 and $E_3$=72; decrease 4.1 and 19.5 times at $E_4$=84 and $E_5$=135.  Survival curves $\ell$ (X, $E_k$) at the intersections and maximal E=295 are presented in Fig. 1b. They yield maximal SLS ~120, 225, 425 years.

The most specific mortality characteristic is its rate q (X, E), i.e. the probability to die at the scaled age X. In the case of restricted heterogeneity, it reduces to mortality values q(X, $E_k$) at the intersections. Sufficiently low mortality, thus low number of deceased, implies significant mortality fluctuations even in humans. Unlike smooth female survivability in 2002 Iceland (total population ~ 500,000) in Figs. 1a, 1b, its mortality rate is stochastically irregular till X~70. For instance, female mortality rate in 2002 Iceland (E=82.3) equals 0.02; 0; 0.0005 at X=35; 36; 37, and 0.06; 0.002; 0.001 at X=58; 59; 60 years; q(X)=0 (i.e. nobody dies) at 19 different ages from 3 till 36 years.  That is why Fig. 1d presents mortality rate in larger and more homogeneous human populations of 1877, 1931, 1998 Switzerland, 1951, 1993 Japan, 1847 Sweden (6) at E $\approx E_1$, $E_2$, $E_4$. All populations manifest large stochastic fluctuations at ages when statistics is poor (e.g., around 10 years of age ~50 girls died in 1993 Japan and 1931 Switzerland, 1-5 in 1998 Switzerland; unreliable data in very old age are omitted). When statistics is reliable, Fig. 1d presents on a semi-logarithmic scale predominantly universal piecewise linear dependence on X. (This implies that in a restricted heterogeneity population with $E=E_k$, mortality rate dependence reduces to the set of Gompertz laws). Its slope is negative at X<10 and positive at X>10. Beyond X=30 years its intersections are close to the universal SLS intersections $E_k$ in Fig. 1c. Exponential mortality rate decrease and increase (at E~82 more than 10,000-fold from X=10 to X=90 years) in Fig. 1d are unprecedented in the destruction rate of any but live systems. They suggest that vital biological defences [e.g., of immune responses (38)] increase in the evolutionary crucial pre-reproductive, and decrease in evolutionary disposable post-reproductive age correspondingly result in exponential decay and proliferation of lethal destructive elements.

Human survivability in Figs. 1a-1b is calculated from mortality rate values in the year of death (like those in Fig. 1d), and in any other way disregards different life histories at different years in different countries. Yet, close mean lifespan values yield close survival curves for $e \approx 20$ in 1860 Iceland and 1773 crop failure Sweden; 32 in 1762 Sweden and Iceland 120 years later; 42 in 1900 Finland and 1847 Iceland; 64 in 1952 and 1922, 83 in 1993 Japan and 2004 Switzerland, despite tuberculosis epidemics in 1890-1940 Finland and pre-1949 Japan. Life tables also disregard immature stages and their different time spans (e.g., a week for drosophilae and up to 17 years in Magucicada larvae) and living conditions, as well as multitude of various factors describing different biology and living conditions of species from anaerobic yeast to human. Yet, all this relatively little affects universality of their survival curves. Figures 1b, 1d allow one to estimate the time of adjustment to the single variable (instantaneous SLS) at, but not significantly quicker than, ~10% of the mean lifespan. Possibility of such express adjustment is related to, and provided by, unusual universal adaptation.

Universal law in Figs. 1a-1d implies that *all its violators perished (or did not emerge) in the previous evolution. Since all evolutionary survivors yield this law, it is the law of universal evolutionary selection of those survivors who rapidly and accurately adapt to the "maximal survival trajectory" $\ell$ (X, E) in any environment.* Such adaptation implies rapid accurate evaluation of the environmental changes, biological response that provides adjustment to the *"maximal survival trajectory"*, and its implementation. This is possible only via programming by universal operating system of non- coding DNA.

Universal law implies that rapid accurate adaptation was vital for survival of diverse species. This uniquely relates it to 5 Major Mass Extinctions- MMEs (96% of marine species perished in the most drastic extinction from 248 to 238 million years ago), which are consistent with reduction of the universal survival trajectory to 5 crossovers in Figs.1c, 1d. MMEs are considered *non- constructive*, but the universal law implies that rapid universal adaptation was *constructively vital* for survival (thus evolved) in MME sudden, rapidly lethal challenges. Genetic programs of MME adaptation depended on the nature of MME, speed of reproduction, etc. This is consistent with 98.5% of non-coding DNA (with no instructions for making proteins, thus "dark matter") in human, 24% in microbe *Rickettsia prowazekii* (43)*, ~10% in Puffer fish, little in yeast; different DNA frequencies of repetitive sequences (29), long range order (44), etc. Universal law of evolutionary selection allows for its experimental study in the few days (i.e. extremely rapid on the evolutionary scale) microbial antibiotic "mini-extinctions"(22-24). If survival in such "mini-extinction" significantly decreases when entire non-coding genome or its specific fragments are removed, then their programming function and MME evolutionary nature are proven. "Mini- extinctions" allow for the study of emerging adaptation programs (possibly including "directed mutations", "behavioral adaptation", etc). A good candidate for such study may be, e.g., the causative agent of epidemic typhus, which has the highest proportion (24%) of non-coding DNA detected so far in a microbial genome (43), and may transform (be programmed?) into antibiotic rifampin and erythromycin resistant (45, 46). Comprehensive microscopic study may suggest the functioning of unusual genomic computer and its operating system. Other test-stone experiments with non-coding DNA removal may include, e.g., mortality adjustment to changing living conditions and development of non-hereditary immunity.

To summarize, non-coding DNA programs non- Mendelian heredity of express adaptation (possibly not only via mortality) according to universal selection law that evolved in major mass extinctions and complements Darwinian evolution.

## Discussion

Heritable individual traits allowed for discovery and study of coding genes. Heritable universal adaptation may be revealed in dynamics of extensively quantified mortality of different species. Such approach is uncommon for geneticists (who are used to "genetic" experiments) and theoretical physicists (who are used to specific models and their theories) alike, but is indispensable in the case of genetic programs.

Extraordinary phylogenetic irregularity of animal lifespans (6-17) is consistent with MME lethal challenges, more related to environmental changes than to species-specific biology. Human MLS is significantly different in different countries even in the same calendar year: 43.2 and 59.7 in 1925 Japan and Iceland; 18.82 and 48.5 in 1882 Iceland and Norway - see Fig. 2a. In contrast, universal express selection proceeds via non-coding DNA programmed adaptation to the unique maximal survival trajectory, which in restricted heterogeneity populations depends only on SLS E and scaled age X and is independent of any other biological and environmental factors. Any adaptation to environment is unique for live systems, being a must for their survival. However, universal express adaptation is unusual. Stochastic mortality implies ultimate inaccuracy (which is the higher, the smaller population size, survivability and mortality rate are) of the universal law- see Fig. 2b (note that ~40 points with zero mortality rate are missing on its semi-logarithmic scale).

Universal law implies "biological relativity" to age transformation X=Fx in any species, thus quantifies certain universality in their biological complexity, as well as applicability of animal models to humans (whose aging population poses a formidable problem in developed countries). Unanticipated evolutionary and genetic nature of mortality may be one of the reasons why leading theoretical and experimental biologists agree: "Aging is the most familiar yet least well-understood aspect of human biology" (40); "Aging is a fundamental, unsolved mystery in biology"(21). Conclusions of different leading experts are controversial. Demographers predict significant increase in human life expectancy within a decade (47). Biogerontologists suggest its proximity to current value in developed countries due to fundamental biological constrains (48). Biologists challenge this paradigm (17).

Perspectives of programmed biological adaptation may hardly be overestimated. An example of hereditary express changes is increase in human brain efficiency. There were only ~10,000 generations of Homo Sapience Sapiens. Presumably, the main ingredients of the most complex biological system- human brain were developed long ago, while recent hereditary changes (related to a minor mass extinction or spontaneous) upgraded its efficiency, and its further evolution rapidly accelerates (30), c.f. also (49).

Consider implications of the universal law for modern species. Slower than adjustment time improvement in living conditions may decrease mortality and "rejuvenate" it (in particular, with resources which are sufficient for the required biological repair) to the values at younger ages. Indeed, within ~6 years of the unification of East and West Germany, mortality in the East converged toward the West's significantly lower levels, especially among people in their 80s and 90s, despite ~45 years of divergent life history (50). Female mortality at 56 years of age in 1998 Switzerland and 1993 Japan decreased correspondingly to its values at 20 in 1931 Switzerland and at 16 years in 1951 Japan (6). Mortality rate of the Norway females born in 1900 was the same at 40 and at 12, and again at 59 and at 17 years of age, when females were 3.5 times younger (6). Dietary restriction initiated in *Drosophila* on day 14, in 3 days restored its mortality at 7 days of age ( 51), when drosophilae were 2.5 times younger. A similar effect was observed in rats (52). The law estimates adjustment time at ~10% percent of the mean lifespan for all animals. More rapid change in mortality, driven by life at previous conditions, may persist compared to the control animals (51, 52).

Consider inherent limitations on the accuracy and validity of the universal law. Living conditions, as well as their rapid change, may also be different at different ages (e.g., contaminated food and water in 1851-1900 England little affected breast-fed infants) compared even to short adjustment time - see, e.g., Fig. 2a. This may yield different population heterogeneity at different ages, and imply universal law violations, e.g., mortality rate oscillations in Fig. 2c. Figure 2d presents mortality rate dependence on age for 2005 Russian males with $E=58.9 \sim E_2$. It is close to universal dependence at $E \sim E_3 \sim 72$ till $X \sim 7$; at $E \sim E_1 \sim 40$ from $X \sim 27$ till $X \sim 70$; and to $E \sim E_2 \sim 60$ beyond $X \sim 80$. Such situation is especially explicit in the scaled survivability curves (Fig. 2e) of genetically identical inbred Drosophilae males in presumably the same shell vials. Their SLSs are $E=31.7$ and 55.2 scaled years for 2x2 (red); 34.7 and 51.8 for 3X3 (green) lines. Mortality of 1X1 line (blue) with E=63.9 was significantly higher till 41 (11 vs. 5% died prior to 18 year of scaled age), and lower thereafter (1.4 vs. 9.7% survived to 90 years) than for the same line with very close E=61.8. Characteristically, scaled survival curves are universal and close for Drosophilae with E=63.9 and 1951 Japanese females with E=62.6. These examples elucidate how deviations from the universal law quantify (hitherto unspecified) living conditions of species as diverse as human and fly.

Mortality of protected species is negligible compared to lethal MME mortality. Thus, universality of the former is a legacy (and possibly also a byproduct) of, which may be linked to, other universal mechanisms crucial for survival in MME. This is consistent with Williams antagonistic pleiotropy and Kirkwood disposable soma theories (40). Non-coding genome programs allow for artificial biological guidance (e.g., life extension, mortality decrease and rejuvenation, arguably even brain upgrade), although possibly at the price of linked universal implications. Indeed, human maximal lifespan increased by mere 1.5% since ancient Rome (where indication of birth and death dates was mandatory on tombstones), while specific mutations increased maximal lifespan 1.6-fold in dwarf Ames mice and 3.6-fold in nematode. Mutated Methuselah nematodes were vital and healthy, yet unfit (53,54,55), thus evolutionary doomed.

MME universal evolutionary dynamics is dominated by survival of non-interacting animals, thus reduces to "master equation" in physics. Once its variables are established (in quantitative experimental study), it yields the MME universal law (which generalizes the derived exact law of universal post- evolutionary dynamics). As a result, unanticipated universality of biological and evolutionary complexity becomes an exact science problem, which allows for unusual verifiable predictions, computer modeling, system biology approach, and further refinements of the law.

The main outstanding problems are biology of vital universal mechanisms; rapid evaluation of and direction to the survival trajectory; most important, genetic programming of biological "navigation". Less universal than MME threats to survival, from ice ages to pandemics and diseases to species-specific challenges, yield spurts in the Gould-Eldridge punctuated equilibrium (56, 57). They may refine the universal law and quantify its universal evolutionary tree (in particular, with branching and singularities).

## Methods

Mortality of protected populations reduces to statistics of non- interacting animals in a given environment. Such problem is tailor-tailed for theoretical study, which always starts with the derivation of the new phenomenological law and analysis of its specifics.

In Figs. 1a-1d all but one population are heterogeneous, thus they demonstrate predominant universality to certain heterogeneity. Denote a homogeneous group by the subscript number G and averaging over all these groups in the population by $< >$. Universality of the probability $\ell$ (X, E) to survive to the scaled age X in the population with the SLS E yields $< \ell$ (X, $E_G$)$> = \ell$ (X, $<E_G>$). Mathematically this implies (58, 59) that all $E_G$ in such heterogeneous population are restricted to a single universal interval $E_k < E_G < E_{k+1}$ ("restricted heterogeneity"); that $\ell$ is universal piecewise linear function of E:

$\ell$ (X,E)=[($E_{k+1}$-E)$\ell_k$(X)+(E-$E_k$)$\ell_{k+1}$(X)]/($E_{k+1}$-$E_k$);   $\ell_k$(X)=$\ell$(X,$E_k$);  k=1,2,…;     (1)

and that rate r=d$\ell$/dE jumps at the universal SLS crossovers $E_k$. Equation (1) reduces all multitude of factors describing population genotypes, phenotypes, environment, living conditions, life history (immature stages included), etc, to species-specific number F, population specific SLS E, and universal functions $\ell_k$(X)of X only. In a general case of arbitrary heterogeneity, Eq. (1) relates $\ell(X,E)$ to the crossover survivabilities $\ell_k(X)$ and their concentrations $c_k$ in the population:

$$\ell(X, E) = \sum c_k \ell_k(X), \quad \sum c_k = 1, \quad \sum c_k E_k = E \qquad (2)$$

According to Eq. (2), mortality rate $q(X,E) = - d\ell(X,E) / \ell(X,E) dX$ equals

$$q(X,E) = [\sum c_k \ell_k(X) q_k(X)] / [\sum c_k \ell_k(X)]; \quad q_k(X) = - d[\ln \ell_k(X)] / dX \qquad (3)$$

Figure 1d implies that

$$q_k(X) = q_{kj} \exp(\rho_{kj} X) \quad \text{if} \quad X_j < X < X_{j+1,} \qquad (4)$$

where $\rho_{k1} < 0$ and $\rho_{kj} > 0$ when $j > 1$. Since $q_k(X)$ is continuous at each $X = X_j$, and $E_k = \int_0^\infty \ell_k(X) dX$, so $q_{kj}$ reduces to the universal numbers $E_k$ and $\rho_{kj}$, $i = 0, 1, \ldots, j$. In the restricted heterogeneity case of Eq. (1), $q_k(X) = q(X, E_k)$ reduces to the set of Gompertz laws. Within stochastic accuracy, Eqs. (2-4) agree with all survival and mortality curves, e.g. in Fig. 2f. Deviations in Figs. 2d, e are related to too rapid (age specific included) changes in living conditions with time t and age X which yielded t and X dependent $c_k$ in Eq. (3).

Equations (1,3) relates E to the value of survivability (for humans in a given calendar year) at, e.g., X= 2 years of age. (By Fig. 1d, the accuracy at X<2 is significantly lower). Thus, at any age adjustment time does not exceed 3 years since conception, i.e., in virtue of universality, 0.1E. So, universal mortality law implies accurate and extremely rapid (yet slower than 0.1E) non-hereditary adjustment to environmental changes.

By Eqs. (3,4),

$$\ell_k(X) = \exp[-\int_0^X q_k(X')dX'] \propto \exp[-q_{kj}/\rho_{kj})\exp(\rho_{kj}X)]$$  (5)

When stochastic fluctuations are sufficiently low (i.e. mortality rate and population size are sufficiently high), piecewise linear dependence of $\log[q_k(X)]$ on X dominates.

In populations whose heterogeneity (higher at younger ages) is not restricted to a single universal E interval, universal linear dependence on the concentrations $c_k$ in Eq. (3) allows one to verify universality of the linear regressions in Fig. 2f. [Note that e=$\int_0^\infty \ell(x)dx \approx 0.5 + \sum_{x=0}^\infty \ell(x)$; $q(x) \approx \ln[\ell(x)/\ell(x+1)]$, where x is the (integer) age from the life table]. This is most explicit in the probability d(X,E) for live newborns to die at the age X: by Eq.(2), d(X,E) is a piecewise linear function of E:

$$d(X, E)=d\ell/dX = \sum c_k d_k(X), \quad \sum c_k=1, \quad \sum c_k E_k=E; \ d_k(X)= d\ell_k/dX.$$  (6)

Demographic data (6) verify(55) this formula.

**Acknowledgments.** I am grateful to Drs. J. Bernstein, G. Falkovich, D. Thaler, A. Vilenkin, I. Korenblit for important comments and suggestions, to I. Kolodnaya for assistance. Financial support from the A. von Humboldt award and R. & J. Meyerhoff chair is appreciated.

In all figures N is the number of animals; x, e and X, E are age, MLS and their scaled values.

Figure 1. Universal law. (a) Dependence of anaerobic yeast (20), nematode (12), human (6) survivability on their relative (i.e. related to the mean lifespan) age x/e. (b) Dependence of survivability on the scaled age X for scaled mean lifespan (from left to right) E~ 20, 33, 42, 64, 84, 135, 295 and (c) to 30 and 85 scaled years vs. scaled mean lifespan E for anaerobic ( 20) and aerobic yeast (42 ), mayfly (11), drosophilae (10), nematode (12, 17), mice (8), human [(6). (Calendar years of, and close to wars, epidemics, etc are not included). In (a-c) full and empty circle, diamond, cross, dash, square, triangle denote anaerobic and aerobic yeast, nematode, mayfly, drosophilae, mice, human. Larger signs indicate higher ages. (d) Universality of female mortality rate (on the semi-logarithmic scale) vs. age for  1993 Japan and 1998 Switzerland (E=82.46 and 82.52 years-only females have so high E); 1951 Japan and 1931 Switzerland (E=62.57 and 63.6); 1847 Sweden and 1877 Switzerland (E=41.91 and 41.52)-  correspondingly lower, middle and upper curves; large, medium and small triangles for Japan, Sweden, Switzerland (6). Solid lines in (c) and (d) are linear regressions.

Figure 2. Universal law, unrestricted heterogeneity, rapid environmental and age specific changes in living conditions. (a) Human mean lifespan e vs. calendar year. The cases in empty circles are considered in other figures. (b) Stochastic fluctuations with age and population size at E~82. Fluctuations are low beyond x=70 in Iceland, 40 in Sweden and Switzerland, 15 in Japan. Zero values (e. g. nineteen from 3 to 36 years in 2002 Iceland) of mortality rate are missing on the semi-logarithmic scale. (c) Deviations from the universality in, e.g., 1773 (crop failure year) Sweden (e=19), 1847 Iceland with small population, 1866 England with contaminated food and water, 1821-25

tuberculoses Japan and 1900 Finland (e~42). (d) Non-universal mortality in 2005 Russia. (e) Survival curves of inbred 1X1 (blue, scaled mean lifespan E=64 and 62), 2X2 (red, E=55 and 32), 3X3 (green, E=52 and 35) Drosophila males (10) in shell vials (larger signs denote larger E), and 1951 Japanese females (black, E=62.6) . (f) Mortality rate (on the semi-logarithmic scale) vs age for mean lifespans e~42, 74, countries with unrestricted heterogeneity cases included. Piecewise linear dependences yield Eqs. (3,4).

**Abbreviations**

MLS –mean life span

SLS – scaled mean life span

MME – major mass extinction

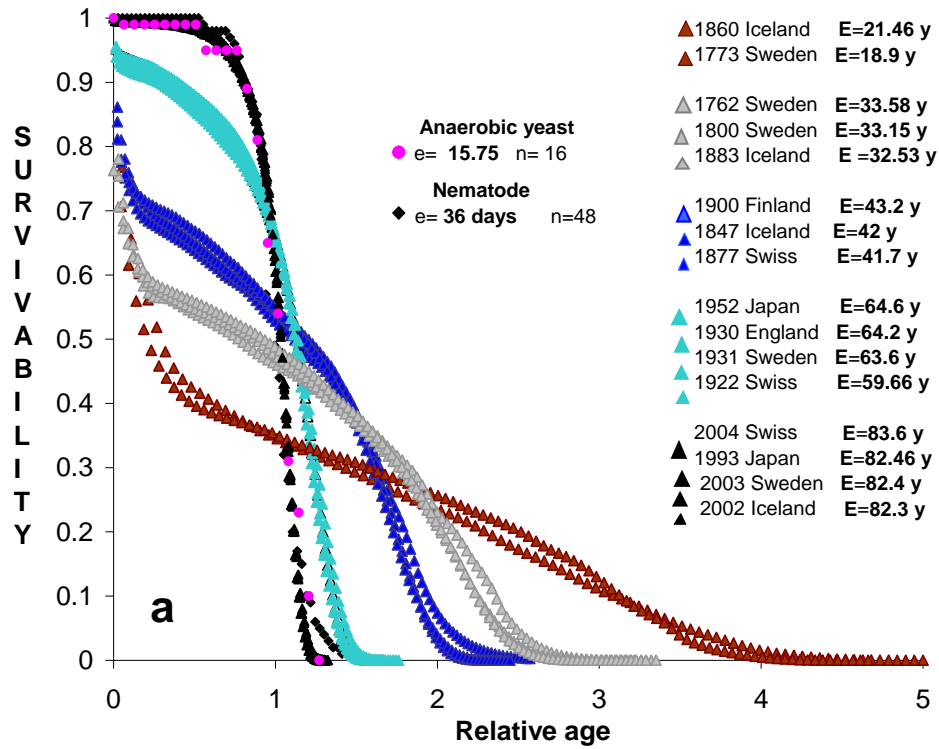

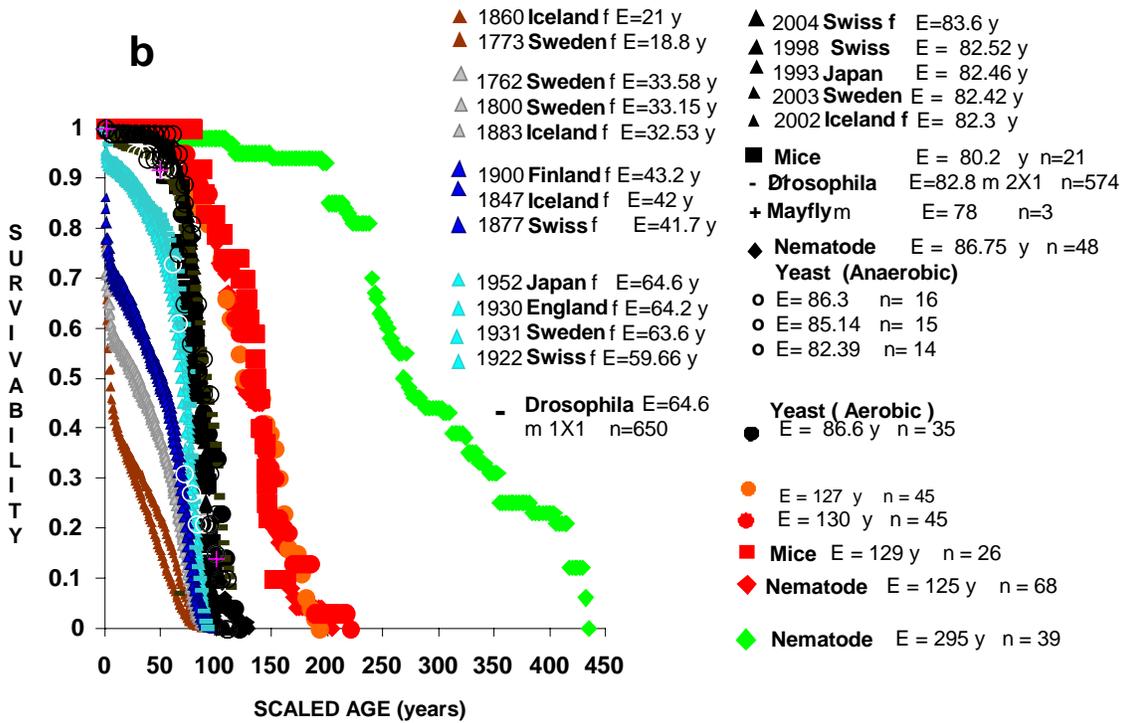

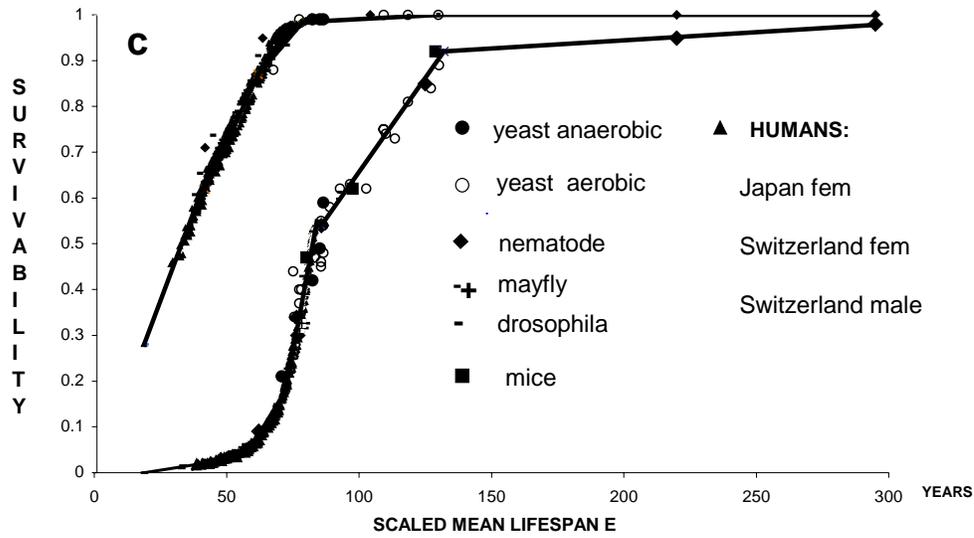

c

S
U
R
V
I
V
A
B
I
L
I
T
Y

- ● yeast anaerobic
- ○ yeast  aerobic
- ◆ nematode
- + mayfly
- - drosophila
- ■ mice

▲ HUMANS:

Japan fem

Switzerland fem

Switzerland male

SCALED MEAN LIFESPAN E    YEARS

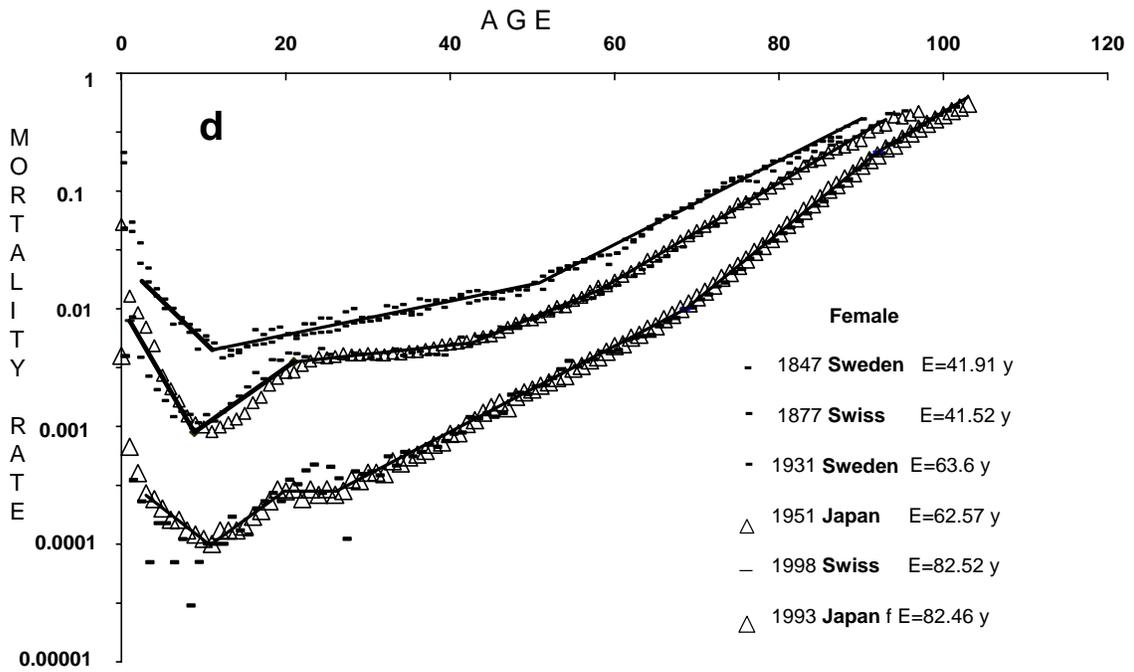

A G E

d

M
O
R
T
A
L
I
T
Y

R
A
T
E

**Female**

- -  1847 **Sweden**  E=41.91 y
- -  1877 **Swiss**   E=41.52 y
- -  1931 **Sweden**  E=63.6 y
- △  1951 **Japan**   E=62.57 y
- -  1998 **Swiss**   E=82.52 y
- △  1993 **Japan** f E=82.46 y

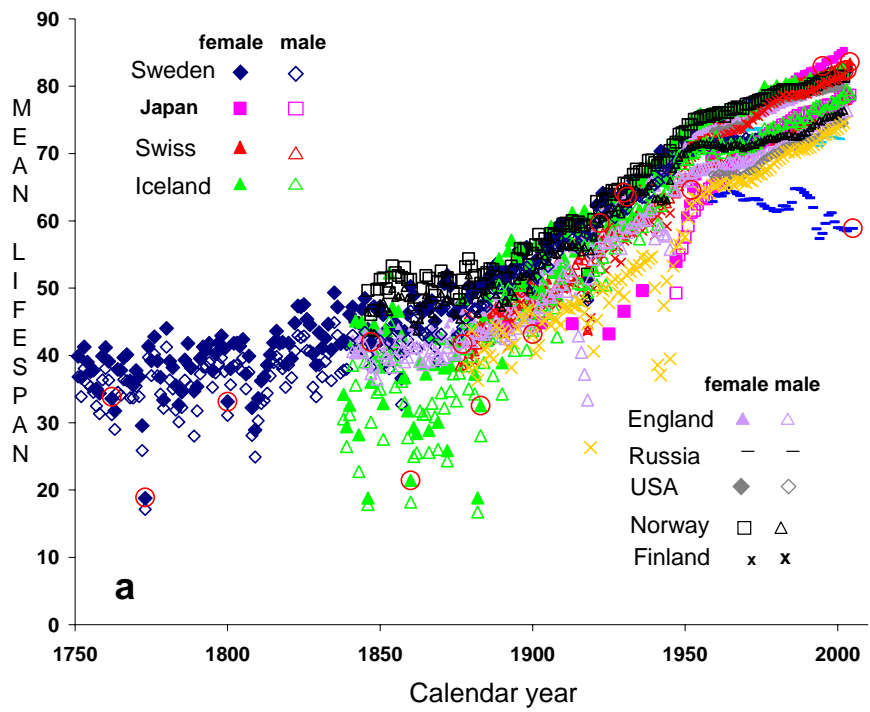

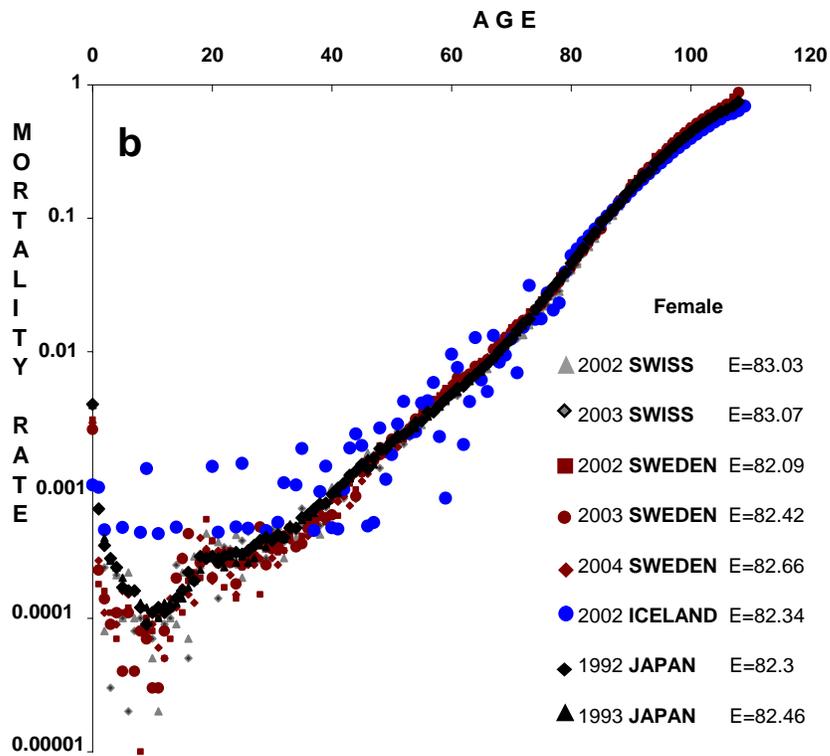

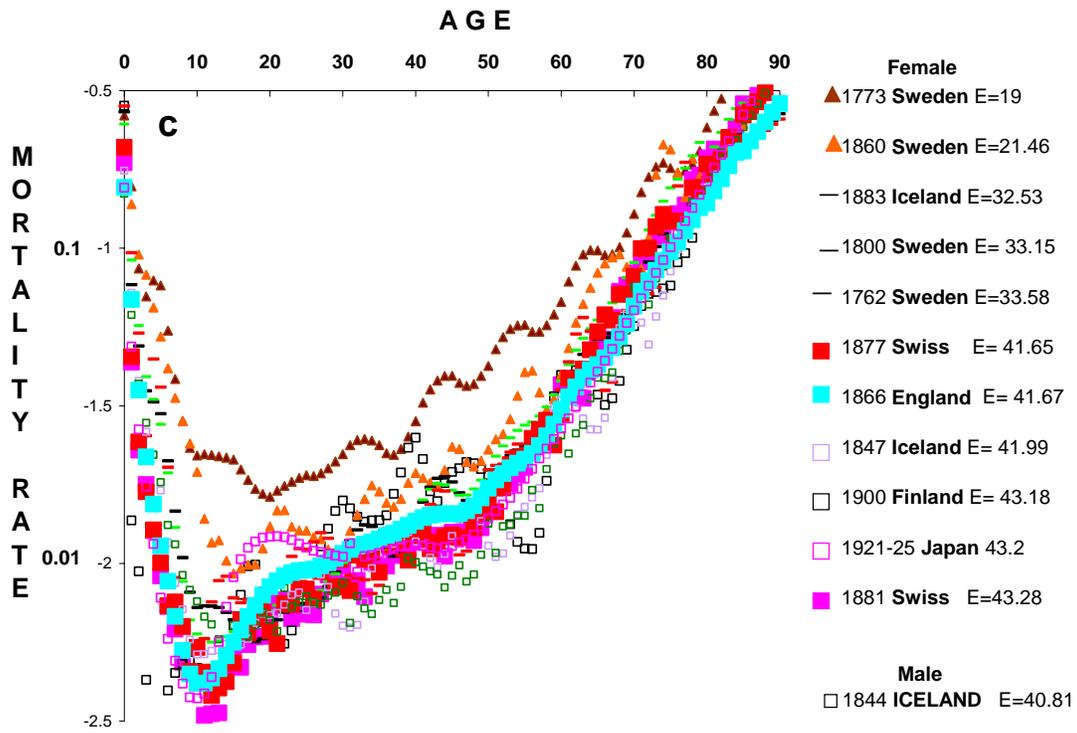

A G E

**Female**
▲ 1773 Sweden E=19
▲ 1860 Sweden E=21.46
— 1883 Iceland E=32.53
— 1800 Sweden E= 33.15
— 1762 Sweden E=33.58
■ 1877 Swiss E = 41.65
■ 1866 England E= 41.67
□ 1847 Iceland E=41.99
□ 1900 Finland E= 43.18
□ 1921-25 Japan 43.2
■ 1881 Swiss E=43.28

**Male**
□ 1844 ICELAND E=40.81

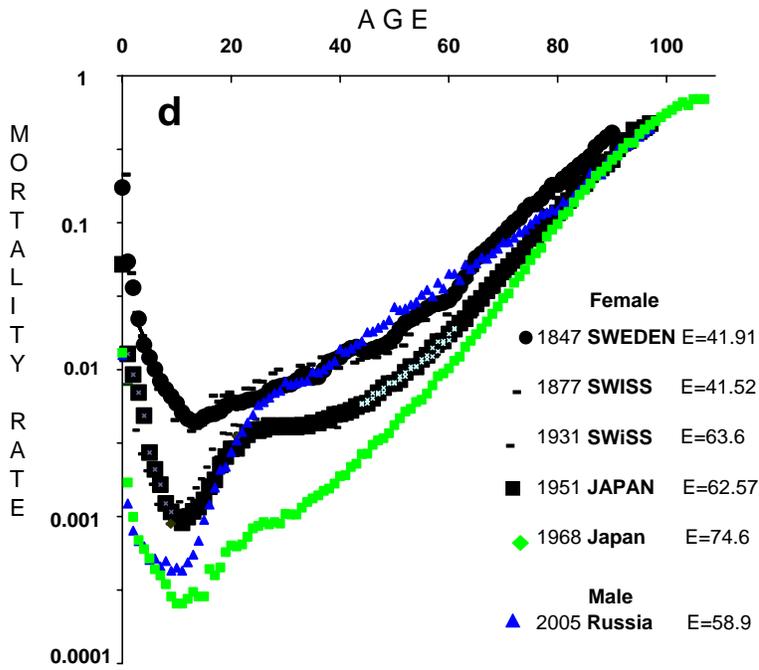

A G E

**Female**
● 1847 SWEDEN E=41.91
▬ 1877 SWISS E=41.52
▬ 1931 SWiSS E=63.6
■ 1951 JAPAN E=62.57
◆ 1968 Japan E=74.6

**Male**
▲ 2005 Russia E=58.9

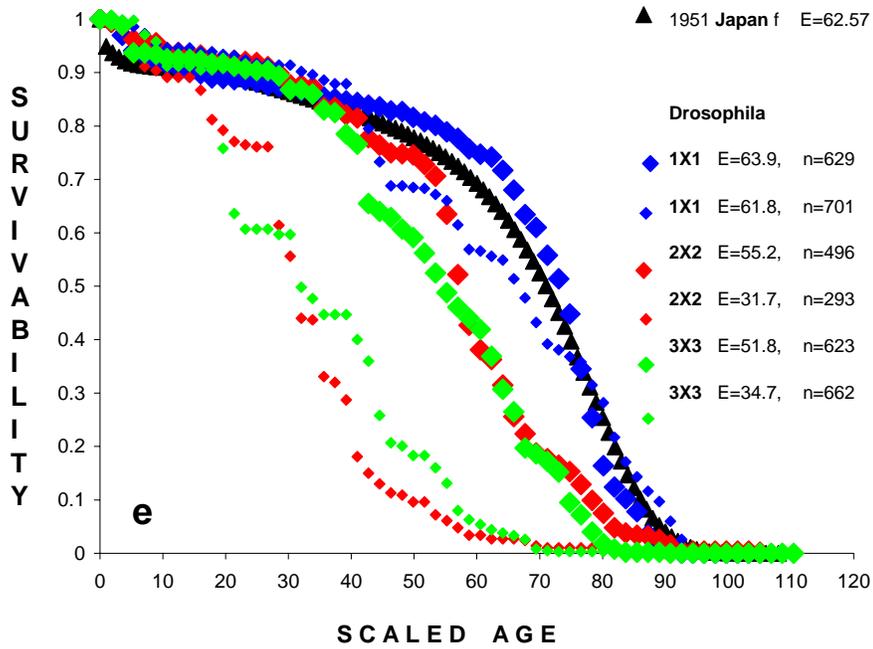

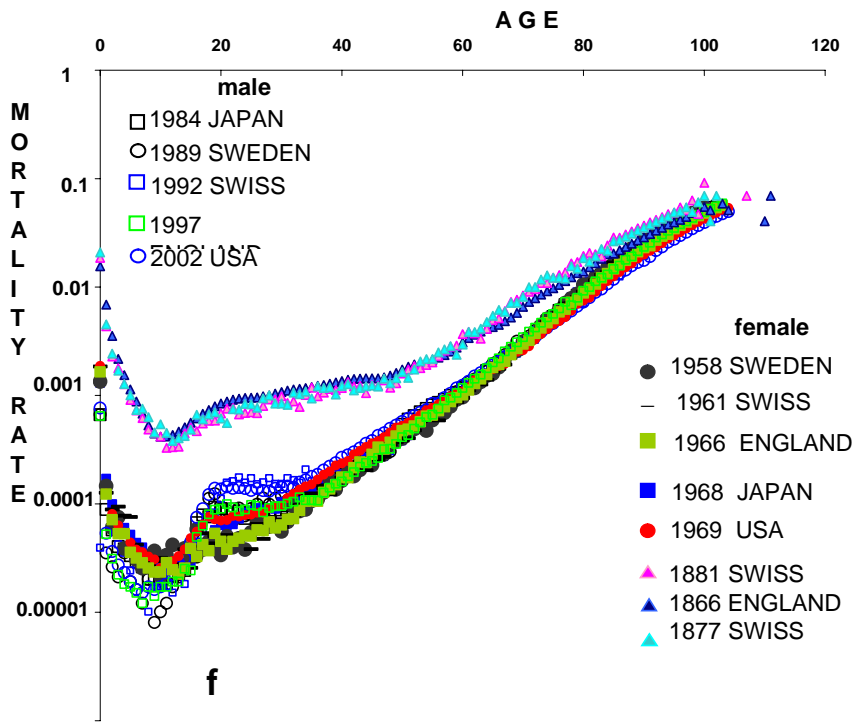